\documentstyle[12pt]{article}

\newcommand{\bb}{\begin{equation}}
\newcommand{\ee}{\end{equation}}
\newcommand{\bqn}{\begin{eqnarray}}
\newcommand{\eqn}{\end{eqnarray}}

\def\no{\nonumber}
\begin{document}


\title{\Large \bf
Boundary terms in the AdS/CFT correspondence for spinor fields}

\author{Marc Henneaux$^*$}
\date{}
\maketitle
\begin{centering}
{\sl
Physique Th\'eorique et Math\'ematique,
Universit\'e Libre de
Bruxelles,\\
Campus Plaine C.P. 231, B--1050 Bruxelles, Belgium\\[1.5ex]

and \\
Centro de Estudios Cient\'\i ficos de Santiago,\\
Casilla 16443, Santiago 9, Chile}

\end{centering}
\vspace{1cm}

\abstract
The requirement that the action be stationary for solutions
of the Dirac equations in anti-de Sitter space
with a definite asymptotic
behaviour is shown to
fix the boundary term (with its coefficient)
that must be added to the standard Dirac action
in the AdS/CFT correspondence for spinor fields.

\vfill
\hfill ULB-TH/99-03

\vspace{1cm}

\noindent{To appear in the Proceedings of the International
Workshop ``ISMP" held in Tbilissi in September of 1998.}

\vspace{.3cm}

\noindent{
$^*$Email: henneaux@ulb.ac.be}

\newpage

\section{Introduction}
\setcounter{equation}{0}

The stationary phase method enables one to give an asymptotic expansion
for the path integral
\bb
Z = \int [D\varphi] \exp \frac{i}{\hbar} S[{\varphi}]
\label{PI}
\ee
of a quantum system in the classical limit $\hbar \rightarrow 0$.
One finds to leading order (and dropping the prefactors)
\bb
Z \sim  \exp\frac{i}{\hbar} S_{cl}
\label{classical}
\ee
where $S_{cl}$ is the value of the action functional $S[{\varphi}]$
evaluated on the classical history, i.e., on the solution of the
classical equations of motion
\bb
\frac{\delta S}{\delta \varphi^i} =0
\ee
fulfilling the given boundary conditions.

It is quite crucial in order for (\ref{classical}) to be
correct that the action $S[{\varphi}]$ be indeed stationary on the
classical path.  This rather trivial observation has far-reaching
consequences on spaces with boundaries since it enables one to
discriminate among the boundary terms that may be added to the
action.  The point is that if a given choice of the
action $S[{\varphi}]$ fulfills $\delta S = 0$ on the classical histories,
it is in general not true that $\delta(S + B_\infty) = 0$ where
$B_\infty= \int \partial_\mu j^\mu d^d x$ is a surface term, since one
may have $\delta B_\infty \not= 0$.
This criterion fixes for instance the form of the conserved charges in 
gauge theories.  It was used in particular in \cite{RT}
to establish that the action for Einstein
gravity must be supplemented by the time-integral of the
ADM mass at infinity in the case of asymptotically flat spaces.

The purpose of this paper is to show that the same criterion 
fixes the form of the boundary term that must be added to 
the standard Dirac action \cite{HS}
in the AdS/CFT correspondence \cite{Maldacena,GKP,Witten}.  Besides
its intrinsic interest, this problem illustrates quite well the
general fact that the action is much more than just a mere mnemonic
device whose only purpose is to concisely summarize
the equations of motion.
Although all Lagrangians ${\cal L}$ that differ by a local divergence
$\partial_\mu j^\mu$ have identical Euler-Lagrange derivatives,
it is only for a special subclass of these Lagrangians
that the Euler-Lagrange equations
are equivalent to the statement that the action $S = \int d^dx {\cal L}$
is stationary in the class of
histories entering the variational principle (i.e.,
$\delta \int d^dx {\cal L}$ is equal to
zero and not a non-vanishing boundary term).

We refer to the literature for background information on
the AdS/CFT correspondence (see \cite{Petersen} for
a recent review) and shall discuss here only the problem
of boundary terms\footnote{The actual talk given in Tbilissi did not
deal with the subject covered in this paper but with the AdS central
charge $c=3l/2G$ that arises in the asymptotic conformal
algebra of $(2+1)$-AdS gravity.}.

\section{Dirac equations on $AdS_{d+1}$}
\setcounter{equation}{0}

We follow \cite{HS} and consider free spinor fields of mass
$m$ on 
(Euclidean) $AdS_{d+1}$.  We take Poincar\'e coordinates $x^\mu =
(x^0, x^i) = (x^0, {\bf x})$ ($i=1, \dots, d)$
such that $AdS_{d+1}$ is the domain $x^0 >0$ with metric
\bb
ds^2 = G_{\mu \nu} dx^{\mu} dx^{\nu} = (x^0)^{-2}
(dx^0 dx^0 + d\tilde{s}^2).
\label{metric}
\ee
In (\ref{metric}),  the $d$-dimensional $d\tilde{s}^2$ is the flat metric
\bb
d\tilde{s}^2 = d {\bf x}  \cdot d {\bf x}.
\ee
In the local Lorentz frame
\bb
e^a_\mu = (x^0)^{-1} \delta^a_\mu
\ee
the Dirac operator
$\not{\!\!D}$ is given by \cite{HS}
\bb
\not{\!\!D} \equiv e_a^{\mu} \Gamma^a(\partial_\mu 
+ \frac{1}{2} \omega^{bc}_\mu \Sigma_{bc})
= x^0 \Gamma^0 \partial_0 + x^0 {\bf \Gamma} \cdot
\nabla - \frac{d}{2} \Gamma^0
\ee
where the matrices $\Gamma^a = (\Gamma^0, \Gamma^i)
= (\Gamma^0,{\bf \Gamma})$
are the flat space gamma-matrices and where the notation
$\partial_\mu = (\partial_0, \partial_i) = (\partial_0, 
\nabla)$ is used.

The Dirac equations read
\bb
(\not{\!\!D} - m) \psi = 0
\ee
for $\psi$ and
\bb
\bar{\psi} (- \overleftarrow{\not{\!\!D}} - m) = 0
\ee
for the conjugate spinor $\bar{\psi}$.
Without loss of generality, we assume $m$ to be positive.

The equations of motion can be solved near the boundary $x^0 = 0$
by the standard Frobenius procedure.  One tries solutions of the
form $(x^0)^\rho \sum_{n=0}^\infty c_n({\bf x}) (x^0)^n$ where 
$c_n({\bf x})$ are $x^0$-independent spinors.  For this series to
solve the equations, $\rho$ must be given by
\bb
\rho = \frac{d}{2} \pm m
\ee
and the first coefficient $c_0({\bf x})$ should be annihilated by
$I - \Gamma^0$ or $I +\Gamma^0$ depending on which value
of $\rho$ is taken.  
There are thus two types of solutions near $x^0 = 0$.  
Solutions of the first type  are
annihilated by the matrix $I + \Gamma^0$ to leading order and
behave asymptotically as
\bb
\psi^{-}(x) = (x^0)^{\frac{d}{2}-m} \psi_0 ({\bf x}) + o\Big((x^0)^{\frac{d}{2}
-m}\Big)
\label{sol0}
\ee
($x^0 \rightarrow 0$) where $\psi_0 ({\bf x})$ is an eigenvector
of $\Gamma^0$ for the eigenvalue $-1$ but is otherwise arbitrary,
\bb
\Gamma^0 \psi_0 ({\bf x}) = - \psi_0 ({\bf x}).
\ee
Solutions of the second type start with higher powers of $x^0$
and are annihilated by $I - \Gamma^0$ to leading order.  They 
behave asymptotically as
\bb
\psi^{+}(x) = (x^0)^{\frac{d}{2}+m} \chi_0 ({\bf x}) + o\Big(
(x^0)^{\frac{d}{2}+m}\Big)
\label{as0}
\ee
($x^0 \rightarrow 0$) where now $\chi_0 ({\bf x})$ is an arbitrary eigenvector
of $\Gamma^0$ for the eigenvalue $+1$,
\bb
\Gamma^0\chi_0 ({\bf x}) = \chi_0 ({\bf x}).
\ee

Similarly, one finds for the conjugate spinors
\begin{eqnarray}
\bar{\psi}^{+}(x) &=& \bar{\psi}_0({\bf x}) (x^0)^{\frac{d}{2}-m} +
o\Big((x^0)^{\frac{d}{2}-m}\Big), \label{sol0'} \\
\bar{\psi}_0({\bf x}) \Gamma^0 &=& \bar{\psi}_0({\bf x})
\end{eqnarray}
and
\begin{eqnarray}
\bar{\psi}^{-}(x) &=& \bar{\chi}_0({\bf x}) (x^0)^{\frac{d}{2}+m} +
o\Big((x^0)^{\frac{d}{2}+m}\Big), \label{as0'} \\
\bar{\chi}_0({\bf x})\Gamma^0 &=& -\bar{\chi}_0({\bf x}).
\end{eqnarray}

One may construct recursively the next terms
in the solutions by using the
Dirac equations. In order to formulate the boundary conditions
below, we note that up to terms of order $o((x^0)^{\frac{d}{2}+m})$,
the solutions  (\ref{sol0}) and (\ref{sol0'}) read
\begin{eqnarray}
\psi^{-}(x) &=& (x^0)^{\frac{d}{2}-m}[I + \sum_{n=1}^p (x^0)^n \alpha_n]
\psi_0 ({\bf x}) + o\Big((x^0)^{\frac{d}{2}+m}\Big), \label{as1} \\
\bar{\psi}^{+}(x) &=& (x^0)^{\frac{d}{2}-m} \bar{\psi}_0({\bf x}) 
[I + \sum_{n=1}^p \overleftarrow{\beta}_n (x^0)^n ]
+ o\Big((x^0)^{\frac{d}{2}+m}\Big)
\label{as2} 
\end{eqnarray}
where $p$ is the biggest integer such that $p \leq 2m$ and where
$\alpha_n$ and $\beta_n$ are definite differential operators
in $d$ dimensions which can be computed from
the Dirac equations (e.g., $\alpha_1 \sim
{\bf \Gamma} \cdot {\nabla}$) and whose explicit form will not be needed in
the sequel\footnote{One may actually need $(\log x^0) \, 
(x^0)^{\frac{d}{2}+m}$-terms
when $m$ is a half-integer but we will not write explicitly
these terms - also determined
by the first terms in the expansion - when they are needed.
The formulas (\ref{onshell}) and (\ref{onshell'}) below
assume $m \not=$ a half-integer.}.

The general solution near $x^0=0$ is obtained by superposition of
$\psi^{-}(x)$ and $\psi^{+}(x)$.  It is determined by the 
boundary data $\psi_0({\bf x})$
and $\chi_0({\bf x})$, 
which may be taken independently
at this stage.  The leading order of the solution
is determined by $\psi_0({\bf x})$, which fulfills
$(I+ \Gamma^0) \psi_0({\bf x}) = 0$.  The other
boundary datum $\chi_0({\bf x})$ is (non trivially)
relevant at lower orders.
Thus, to leading order,
the general solution of the Dirac equation near $x^0 = 0$ 
belongs to the eigenspace of $\Gamma^0$
with eigenvalue $-1$.  The next order belongs to the
eigenspace with eigenvalue $+1$.  
This ``peeling off" property of Dirac fields is
reminiscent of what happens for Rarita-Schwinger
fields \cite{HT}.

If one demands that the solution
be well-behaved in the volume of anti-de Sitter space
$AdS_{d+1}$  up to $x^0  = \infty$ (which consists of  
a {\em single} boundary point since the metric along
${\bf x}$ vanishes in the limit $x^0 \rightarrow
\infty$ \cite{Witten,Petersen}\footnote{This point
is the compactification point at infinity of $x^0=0$ \cite{Witten}.
We shall assume that the data $\psi_0({\bf x})$ and $\bar{\psi}_0
({\bf x})$ can be Fourier-transformed in ${\bf x}$ (e.g., are
of compact support) and so vanish for $\vert {\bf x} \vert
\rightarrow \infty$.
For regularity, the solution must then vanish for $x^0 \rightarrow
\infty$.}), one
finds that $\chi_0 ({\bf x})$ and $\psi_0({\bf x})$ cannot be
taken independently.  Rather, they must be
related as
\bb
\chi_0 ({\bf k}) = - i \frac{{\bf k} \cdot {\bf \Gamma}}{k}
\frac{k^{2m} 2^{-2m} \Gamma(\frac{1}{2}-m)}
{\Gamma(m+\frac{1}{2})} \psi_0({\bf k})
\label{onshell}
\ee
where $\chi_0 ({\bf k})$ and $\psi_0({\bf k})$ are the 
respective Fourier transforms of 
$\chi_0 ({\bf x})$ and $\psi_0({\bf x})$, and
where $k = \sqrt{{\bf k}^2}$.  One must similarly
impose
\bb
\bar{\chi}_0 ({\bf k}) = i \bar{\psi}_0({\bf k})
\frac{{\bf k} \cdot {\bf \Gamma}}{k}
\frac{k^{2m} 2^{-2m} \Gamma(\frac{1}{2}-m)}
{\Gamma(m+\frac{1}{2})}
\label{onshell'}
\ee
for the conjugate spinors.

These relations follow from the work of \cite{HS,MV}, where the
general solution that vanishes for $x^0 \rightarrow
\infty$ is explicitly constructed.
One finds 
\bb
\psi(x) = (x^0)^{\frac{d+1}{2}}
\int  \frac{d^d {\bf k}}{(2 \pi)^d} e^{-i {\bf k} \cdot {\bf x}}
\big[A({\bf k}) + B({\bf k}) \big]
\label{sol1}
\ee
where
\bb
A({\bf k}) = -i \frac{{\bf k}\cdot {\bf \Gamma}}{k} K_{m-\frac{1}{2}} (kx^0)
\frac{k^{m+\frac{1}{2}}\psi_0({\bf k})}{\Gamma(m+\frac{1}{2}) 2^{m-\frac{1}{2}}}
\ee
and
\bb
B({\bf k})= K_{m+\frac{1}{2}} (kx^0)
\frac{k^{m+\frac{1}{2}}
\psi_0({\bf k})}{\Gamma(m+\frac{1}{2}) 2^{m-\frac{1}{2}}}.
\ee
Here $K_\nu(z)$ is the modified Bessel function which
vanishes as $z \rightarrow \infty$. 
Similarly, one obtains \cite{MV}
\bb
\bar{\psi}(x) = (x^0)^{\frac{d+1}{2}}
\int  \frac{d^d {\bf k}}{(2 \pi)^d} e^{-i {\bf k} \cdot {\bf x}}
\big[\bar{A}({\bf k}) + \bar{B}({\bf k}) \big]
\label{sol2}
\ee
with
\bb
\bar{A}({\bf k}) = i \frac{\bar{\psi}_0({\bf k})k^{m+\frac{1}{2}}}
{\Gamma(m+\frac{1}{2}) 2^{m-\frac{1}{2}}}
\frac{{\bf k}\cdot {\bf \Gamma}}{k} K_{m-\frac{1}{2}} (kx^0)
\ee
and
\bb
\bar{B}({\bf k}) = \frac{\bar{\psi}_0({\bf k})k^{m+\frac{1}{2}}}
{\Gamma(m+\frac{1}{2}) 2^{m-\frac{1}{2}}}
K_{m+\frac{1}{2}} (kx^0)
\ee
Using the small argument expansion of the Bessel functions,
\bb
K_\nu(z) = 2^{\nu -1} \Gamma(\nu) (z)^{-\nu} [1 + \dots]
- 2^{-\nu -1} \frac{\Gamma(1-\nu)}{\nu} (z)^\nu [1 + \dots]
\ee
where the dots denote positive integer powers of $(z)^2$,
one can verify that the solutions behave as
indicated above for 
$x^0 \rightarrow 0$, and that $\chi_0({\bf x})$ and
$\bar{\chi}_0({\bf x})$ are related to $\psi_0({\bf x})$
and $\bar{\psi}_0({\bf x})$ as written in (\ref{onshell}) and
(\ref{onshell'}).
The leading term in $\psi^- (x)$ comes from
the first term in the expansion of $B({\bf k})$; the leading term
in $\psi^+(x)$ comes from the second term in
the expansion of $A({\bf k})$ (coefficient of $(x^0)^{m-\frac{1}{2}}$).

It follows that the general solution of the Dirac
equations in the whole of AdS
is determined by a single
spinor field annihilated by $1+\Gamma_0$ on the boundary \cite{HS,LR}.  Half
of the boundary data are expressed in terms of the other
half if the solution is to be a solution
everywhere.
It should be stressed, however, that (\ref{onshell}) and (\ref{onshell'})
are on-shell
relations valid for solutions of the classical equations
of motion but not for all competing histories considered
in the variational principle. In fact, one
can easily construct off-shell configurations which 
are regular in the whole of $AdS_{d+1}$ and which involve
independent $\psi^-$ and $\psi^+$ in the vicinity of
$x^0 =0$.

\section{Variational principle}
\setcounter{equation}{0}

Because the Dirac equations contain only first order
derivatives, one cannot fix simultaneously all the components
of $\psi$ and $\bar{\psi}$ at $x^0 = 0$.  This would be like
fixing both $q$'s and $p$'s at the boundary, and this is not
permissible.  Rather, one can only fix a complete set of
``commuting" variables, e.g. the $q$'s or the $p$'s.  
For the AdS/CFT correspondence, it
is appropriate to fix the components $\psi_0$ and $\bar{\psi}_0$,
which are the sources for the Green functions in $d$ dimensions \cite{HS}
and to leave $\chi_0({\bf x})$ and $\bar{\chi}_0({\bf x})$ free to
vary.
We shall thus consider in the variational principle all configurations which
take the form
\begin{eqnarray}
\psi(x) &=& \psi^- (x) + \psi^+(x), \\
\bar{\psi}(x) &=& \bar{\psi}^+ (x) + \bar{\psi}^-(x)
\end{eqnarray}
where $\psi^- (x)$ and $\bar{\psi}^+(x)$ behave asymptotically
($x^0 \rightarrow 0$) as in (\ref{as1}) and (\ref{as2}) with
given $\psi_0({\bf x})$ and $\bar{\psi}_0({\bf x})$, while
$\psi^+(x)$ and $\bar{\psi}^-(x)$
behave asymptotically as (\ref{as0}) and (\ref{as0'}) with
coefficients $\chi_0({\bf x})$ and $\bar{\chi}_0({\bf x})$
that are free to vary.  The $o\Big((x^0)^{\frac{d}{2} +m}\Big)$ terms in
$\psi(x)$ and $\bar{\psi}(x)$ are of course also allowed to vary
and need not be such that the histories $\psi(x)$ and
$\bar{\psi}(x)$ are solutions of the Dirac equations.

So, one does not (and cannot) impose any relationship between
$\chi_0({\bf x})$ and $\psi_0({\bf x})$ or $\bar{\chi}_0({\bf x})$
and $\bar{\psi}_0({\bf x})$ in the variational principle.  The
relations (\ref{onshell}) and (\ref{onshell'}) emerge on-shell.

If one varies the standard Dirac action
\bb
S_D = \int_{AdS} d^{d+1} x \sqrt{G} 
\bar{\psi} \Big[ \frac{1}{2}\big( 
\overrightarrow{\not{\! \! D}} - \overleftarrow{\not{\! \! D}}\big)
-  m \Big]\psi
\ee
with respect to $\psi$ and $\bar{\psi}$ in the class of field configurations 
just defined, one finds, keeping all
surface terms,
\bb
\delta S_D = B_\infty + \hbox{ {\small {\em terms that vanish when the
Dirac equations hold}}}
\ee
where $B_\infty$ is given by
\bb
B_\infty = - \frac{1}{2} \int d^d {\bf x}
\big[\bar{\psi}_0({\bf x}) \delta \chi_0({\bf x})
+ \delta \bar{\chi}_0({\bf x}) \psi_0({\bf x})\big]
\label{variation}
\ee
(recall that $\sqrt{^{(d+1)}\!G} = (x^0)^{-d-1}$).
This term is itself the variation of a surface term at infinity,
\bb
B_\infty = - \delta C_\infty
\ee
with
\bb
C_\infty = \frac{1}{2} \int d^d {\bf x} \big[ \bar{\psi}_0({\bf x})
\chi_0({\bf x}) + \bar{\chi}_0({\bf x}) \psi_0({\bf x}) \big]
\label{Cinfini}
\ee
since $\delta \bar{\psi}_0({\bf x}) = 0$ and 
$\delta \psi_0({\bf x}) = 0$.  

Because $B_\infty \not= 0$, the action $S_D$ is not
stationary on the Dirac solutions in the class of
competing histories defined by the boundary conditions
that $\psi_0$ and $\bar{\psi}_0$ are fixed while $\chi_0$
and $\bar{\chi}_0$ are free to vary.
However,
the ``improved action" obtained by adding $C_\infty$
to $S_D$,
\bb
S= S_D + C_\infty
\ee
is such that $\delta S = 0$ on-shell.
It is therefore $S$ (and not $S_D$) that must be used in the
AdS/CFT correspondence.

Three comments are in order:
\begin{enumerate}
\item Since $\bar{\psi}_0({\bf x})$ and $\psi_0({\bf x})$ are fixed,
one may add any function of these variables to $S$ without changing the
property that $\delta S = 0$.  Such terms correspond to
``phase transformations".  However, if one requires that the surface term
(i) is local; (ii) does not contain derivatives (because the bulk
part is first-order); and (iii) preserves the AdS symmetry, then, 
$(\ref{Cinfini})$ appears to be the only choice ($\bar{\psi}_0 
\psi_0 = 0$).  Note in particular that the coefficient of
$C_\infty$ in (\ref{Cinfini}) is completely determined.
\item Up to terms which give delta-function contact terms in the
correlators and which are thus unimportant \cite{FMMR}
(even though they may contain powers of $\epsilon$ that
blow up as $\epsilon \rightarrow 0$), one may 
rewrite the
surface term $(\ref{Cinfini})$ as
\bb
C_\infty = \lim_{\epsilon \rightarrow 0} \frac{1}{2}
\int_{M_\epsilon} d^d{\bf x} \sqrt{^{(d)} G_\epsilon}
\bar{\psi}\psi
\ee
where $M_\epsilon$ is a $d$-dimensional surface that approaches
the boundary of $AdS_{d+1}$ as $\epsilon$ goes to zero,
and where $^{(d)} G_\epsilon$ is the determinant of the
induced metric on $M_\epsilon$.  This agrees with \cite{HS}.
\item The kinetic term in the action (with $x^0$ being the ``time") shows that
$\bar{\psi}_0({\bf x})$
and $\chi_0({\bf x})$ (respectively, $\bar{\chi}_0({\bf x})$
and $\psi_0({\bf x})$) form  canonically conjugate pairs.  So, one is indeed
fixing the $q$'s and leaving the $p$'s free to vary at the boundary.
\end{enumerate}

\section{Correlation functions}
\setcounter{equation}{0}

We can now compute the correlation functions, following closely
\cite{HS,MV}.  To that end,
one needs the value $S_{cl}$ of the classical
action.  As stressed
in \cite{HS}, the standard Dirac action vanishes on-shell and the
whole contribution comes from the surface term.  Replacing in $C_\infty$
the spinors $\chi_0({\bf x})$ and 
$\bar{\chi}_0({\bf x})$ by their on-shell values
given by (\ref{onshell}) and (\ref{onshell'}) yields
\begin{eqnarray}
S_{cl} &=& \frac{1}{2} \int \frac{d^d {\bf k}}{(2 \pi)^d}
\big[ \bar{\psi}_0({\bf - k}) \chi_0({\bf k}) +
\bar{\chi}_0({\bf k}) \psi_0({\bf - k}) \big] \no \\
&=& i \frac{2^{-2m}\Gamma(\frac{1}{2}-m)}{\Gamma(m+\frac{1}{2})}
\int \frac{d^d {\bf k}}{(2 \pi)^d}
\bar{\psi}_0({\bf k}) \frac{{\bf k} \cdot {\bf \Gamma}}{k} k^{2m}
\psi_0({\bf - k})
\end{eqnarray}
If one rewrite the integral in position space, one gets
\bb
S_{cl} = - \frac{\Gamma(\frac{d+1}{2}+m)}{\pi^{\frac{d}{2}} 
\Gamma(m + \frac{1}{2})} \int d^d {\bf x} d^d {\bf y} \bar{\psi}_0 ({\bf x})
\frac{{\bf \Gamma} \cdot ({\bf x} - {\bf y})}{\vert {\bf x}
- {\bf y} \vert^{d+2m+1}} \psi_0({\bf y})
\ee
which shows that the $2$-point function is
\bb
\Omega({\bf x}, {\bf y}) \sim \frac{{\bf \Gamma} \cdot ({\bf x} - {\bf y})}{\vert {\bf x}
- {\bf y} \vert^{d+2m+1}}
\ee
as shown in \cite{HS,MV} and in agreement with the CFT on the boundary.

\section{Conclusions}
In this paper, we have shown that the surface term needed
in the AdS/CFT correspondence for spinor fields can be understood
from the requirement that the variational principle has a solution
in the relevant class of field histories.  This class of field
histories is characterized by the boundary data at $x^0 =0$:
one fixes the components of the spinor fields that
are annihilated by
$I + \Gamma^0$ but not the components with opposite
``chirality".  These components come with a higher power of
$x^0$ but play a non trivial role because they do contribute to the
boundary terms.  They are unrelated off-shell to the prescribed
components of $\psi$ and are free to vary in the variational principle,
although they become functions of these prescribed components when the
Dirac equations hold everywhere.

\noindent
{\em Note added}: 
A different (but equivalent) justification of the boundary term has been given
recently in \cite{AF}. It is based on the Hamiltonian formalism with
$x^0$ viewed as the evolution parameter.  It is equivalent to the 
requirement that the classical path be a true stationary point, since
this imposes the form $p \dot{q}$ (rather than $-\dot{p} q$, say)
to the kinetic term in the Hamiltonian action in the coordinate
representation where the $q$'s are given at the
boundary \cite{Lanczos}.  I am grateful to Kostas
Sfetsos for pointing out reference \cite{AF} to me.

\section*{Acknowledgements}
The author is grateful to the organizers of the ISMP meeting
for their wonderful hospitality in Tbilissi.
This work has been partly supported by the ``Actions de
Recherche Concert\'ees" of the ``Direction de la Recherche
Scientifique - Communaut\'e Fran\c{c}aise de Belgique" and by
IISN - Belgium (convention 4.4505.86).
Support from Proyectos FONDECYT 1970151 and 7960001 (Chile)
is also gratefully acknowledged.

\end{document}